%% file: gershon_sub.tex
\documentclass[12pt]{article}
\usepackage{epsfig,graphicx}
\usepackage{fpcp04}

\input fpcp04-macro

\newcommand{\model}{\ensuremath{\mathrm{(model)}}\xspace}

\begin{document}

\Title{
  Hot Topics from Belle
}
\bigskip

\label{GershonStart}

\author{
  Tim Gershon
  \index{Gershon, T.} 
}

\address{
  KEK, \\
  Tsukuba, Ibaraki, 305-0801, Japan \\
}

\makeauthor
\abstracts{
  The most precise current determination 
  of the Unitarity Triangle angle $\phi_3$ is presented.
  It is obtained using the decay $B^+ \to D^{(*)} K^+$,
  with Dalitz plot analysis of the subsequent $D \to K_S \pi^+\pi^-$ decay.
  The result is 
  $\phi_3 = 68^\circ \,^{+14^\circ}_{-15^\circ} \stat \pm 13^\circ \syst \pm 11^\circ \model$.  
}

\section{Introduction}

The year 2004 has seen a great number of important new results from Belle.
Studies of time-dependent asymmetries in $B \to \pi^+\pi^-$
revealed the second observation of $CP$ violation,
and the first evidence for direct $CP$ violation, 
in the $B$ system~\cite{bevan}.
More recently, further evidence for direct $CP$ violation has been found
in $B \to K^-\pi^+$ decays; this effect is also seen by BaBar~\cite{graziani}.
The decay $B \to \pi^0\pi^0$ is now observed with more than 5$\sigma$~\cite{belle_pi0pi0},
and evidence for $B \to \rho^0\pi^0$ has been seen~\cite{belle_rho0pi0}.
Studies of new particles observed at Belle have stimulated
theoretical interest~\cite{skchoi},
and further excitement has been aroused by new results on the
time-dependent asymmetries in $b \to s$ penguin transitions~\cite{belle:btos};
new results on $B \to K_S K_S K_S$ and $B \to K_S \pi^0 \gamma$ 
were shown for the first time in FPCP2004~\cite{hazumi}.
To complete this brief selection of 
the many other new results from Belle~\cite{belle_conf},
the first studies of the forward-backward asymmetry in $B \to K^* l^+l^-$
have been performed~\cite{yjkwon}.

This plethora of new results is made possible by the excellent 
performance of the KEKB accelerator.
KEKB is an asymmetric-energy $e^+e^-$ ($3.5 \ {\rm GeV}$ on $8 \ {\rm GeV}$) 
collider~\cite{kekb}, operating at the $\Upsilon(4S)$ resonance 
($\sqrt{s}=10.58 \ {\rm GeV}$) with a peak luminosity that exceeds
$1.3 \times 10^{34}~{\rm cm}^{-2}{\rm s}^{-1}$.
At the time of FPCP2004, the total integrated luminosity was
approaching $300 \ {\rm fb}^{-1}$ 
(at the time of writing it is approaching $330 \ {\rm fb}^{-1}$),
however the results presented here use the data sample of 
$253 \ {\rm fb}^{-1}$ accumulated by summer 2004, 
corresponding to approximately $275 \times 10^6$ $B\bar{B}$ pairs.

The Belle detector is a large-solid-angle magnetic spectrometer that
consists of a three-layer silicon vertex detector (SVD),
a 50-layer central drift chamber (CDC), 
an array of aerogel threshold \v{C}erenkov counters (ACC), 
a barrel-like arrangement of time-of-flight scintillation counters (TOF), 
and an electromagnetic calorimeter comprised of CsI(Tl) crystals (ECL) 
located inside a superconducting solenoid coil that provides a 
$1.5 \ {\rm T}$ magnetic field.  
An iron flux-return located outside of the coil is instrumented 
to detect $K_L^0$ mesons and to identify muons (KLM).  
The detector is described in detail elsewhere~\cite{belle}.
Two different inner detector configurations were used. 
For the first sample of 152 million $B\bar{B}$ pairs, 
a 2.0 cm radius beampipe and a 3-layer silicon vertex detector were used;
for the latter 123 million $B\bar{B}$ pairs,
a 1.5 cm radius beampipe, a 4-layer silicon detector
and a small-cell inner drift chamber were used~\cite{ushiroda}.

\section{Principles of $\phi_3$ Extraction from $B^+ \to D^{(*)} K^+$}

The main objective of the $B$ factories is 
to make precise measurements of the elements of the 
Cabibbo-Kobayashi-Maskawa (CKM) quark mixing matrix~\cite{ckm},
As is well known, the unitarity of the CKM matrix 
results in the expression
\begin{equation}
  \label{eq:cp_uta:ut}
  V_{ud}V_{ub}^* + V_{cd}V_{cb}^* + V_{td}V_{tb}^* = 0,
\end{equation}
which can be represented as a triangle in the complex plane
(the so-called Unitarity Triangle).
Two popular (and various other less popular) sets of names 
for the angles of this triangle are used,
\begin{equation}
  \label{eq:cp_uta:abc}
  \phi_1 \equiv \beta = \arg\left[ - \frac{V_{cd}V_{cb}^*}{V_{td}V_{tb}^*} \right],
  \hspace{0.5cm}
  \phi_2 \equiv \alpha = \arg\left[ - \frac{V_{td}V_{tb}^*}{V_{ud}V_{ub}^*} \right],
  \hspace{0.5cm}
  \phi_3 \equiv \gamma = \arg\left[ - \frac{V_{ud}V_{ub}^*}{V_{cd}V_{cb}^*} \right].
\end{equation}

It has long been known that decays of the type $B \to DK$
are sensitive to $\phi_3$~\cite{bigisanda},
and various methods using different final states of the neutral $D$ meson
have been proposed~\cite{phi3_methods}.
These are based on two key observations:
neutral $D^0$ and $\bar{D}^0$ mesons can decay to a common final state, 
and the decay $B^+ \to D^{(*)} K^+$ can produce neutral $D$ mesons 
of both flavours
via $\bar{b} \to \bar{c}u\bar{s}$ and $\bar{b} \to \bar{u}c\bar{s}$ transitions,
with a relative phase $\theta_+$ that depends on 
both strong and weak phase differences between the two amplitudes,
$\theta_+ = \delta + \phi_3$.
For the charge conjugate mode, the relative phase is $\theta_- = \delta - \phi_3$. 
The size of possible $CP$ violation effects depends on the 
ratio of the magnitudes of the interfering amplitudes, which we denote by $r$.

Recently, three body final states common to $D^0$ and $\bar{D}^0$, 
such as $K_S\pi^+\pi^-$~\cite{ggsz}, 
were suggested as promising modes for the extraction of $\phi_3$. 
The principle of the technique is as follows
(for more details see~\cite{zupan}).
We define the amplitude for $\bar{D}^0 \to K_S\pi^+\pi^-$ 
at a point on the Dalitz plot given by $(m_+^2,m_-^2)$ to be $f(m_+^2,m_-^2)$,
where $m_+^2$ and $m_-^2$ are the squared invariant masses of 
$K_S \pi^+$ and $K_S \pi^-$, respectively.
Assuming $CP$ invariance in $D$ meson decay,
the corresponding amplitude for $D^0$ decay is then $f(m_-^2,m_+^2)$.
The amplitude for $B^+ \to DK^+$ can then be written as 
$f(m_+^2,m_-^2) + r e^{i( \delta + \phi_3 )} f(m_-^2,m_+^2)$,
while that for $B^- \to DK^-$ is 
$f(m_-^2,m_+^2) + r e^{i( \delta - \phi_3 )} f(m_+^2,m_-^2)$.
Thus, once the $D$ decay model $f$ is fixed,
the values of $r$, $\phi_3$ and $\delta$ can be determined by 
simultaneously fitting data from $B^+$ and $B^-$ decays.

The same arguments also apply to similar decays such as $B^+ \to D^* K^+$.
Some care must be taken, 
since for each $B$ decay mode the values of $r$ and $\delta$ can be different.
It has recently been pointed out that for $B^+ \to D^* K^+$ 
there is an effective shift of $\pi$ in the value of $\delta$ 
between the cases that the decays 
$D^* \to D\pi^0$ and $D^* \to D\gamma$ are used~\cite{bg}.

The $D$ decay distribution can be measured at $B$ factories
using the large samples of flavour tagged $D$ mesons produced 
in $D^+ \to D^0 \pi^+$ (and charge conjugate) decays 
following continuum $e^+e^- \to c\bar{c}$ reactions.
However only $\left| f(m_+^2,m_-^2) \right|^2$ can be measured 
resulting in some uncertainty due to 
the model assumed to describe the phase variation across the Dalitz plot.

This technique was first implemented by Belle,
using a $140 \ {\rm fb}^{-1}$ data sample, to obtain
$\phi_3 = 77^\circ\;^{+17^\circ}_{-19^\circ} \stat \pm 13^\circ \syst \pm 11^\circ \model$~\cite{belle_phi3}.
Recently the same method has been used by BaBar, 
with a data sample of $211 \times 10^6$ $B\bar{B}$ pairs, to obtain  
$\phi_3 = 88^\circ \pm 41^\circ \stat \pm 19^\circ \syst \pm 11^\circ \model$~\cite{babar_phi3}.

The results presented below are preliminary updates of the previous 
Belle analysis. More details can be found in~\cite{belle_new}.

\section{Dalitz analysis}

\subsection{Extraction of $D$ Decay Model}

\begin{figure}[htb]
  \begin{center}
    \epsfig{width=0.25\textwidth,file=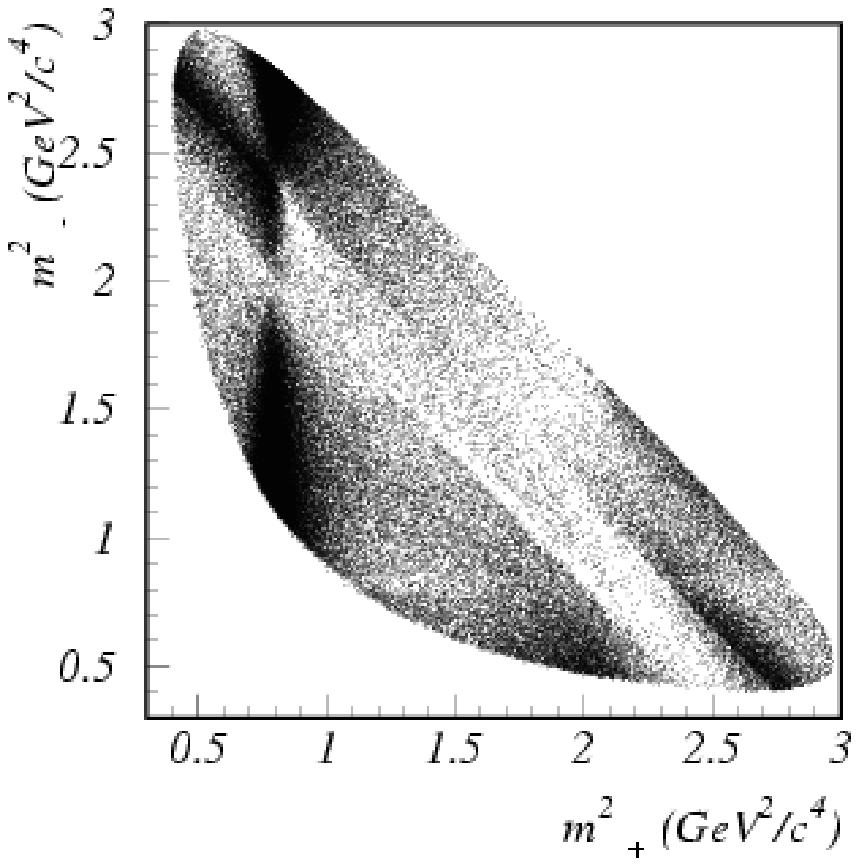}    
    \epsfig{width=0.23\textwidth,file=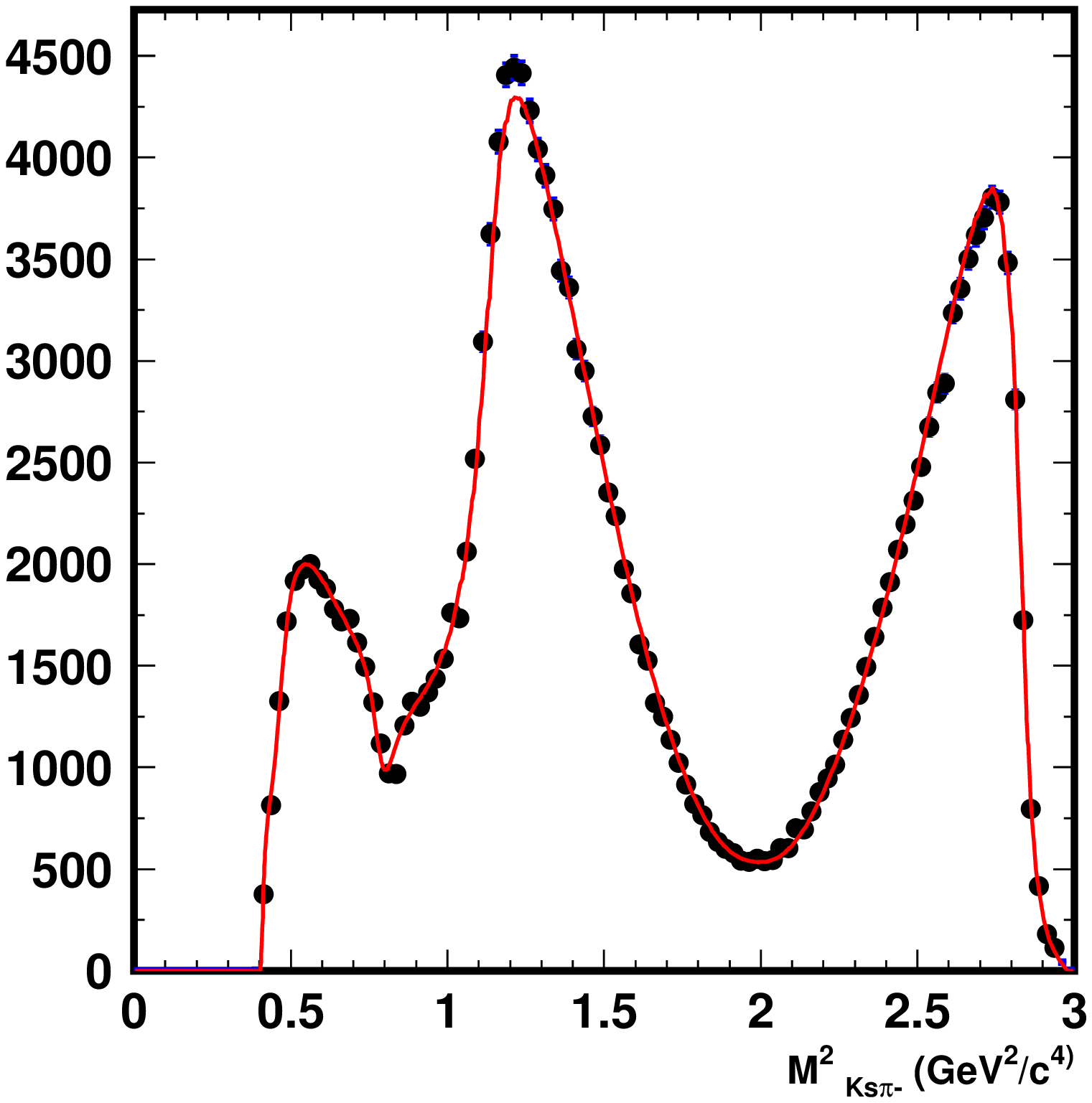}
    \epsfig{width=0.23\textwidth,file=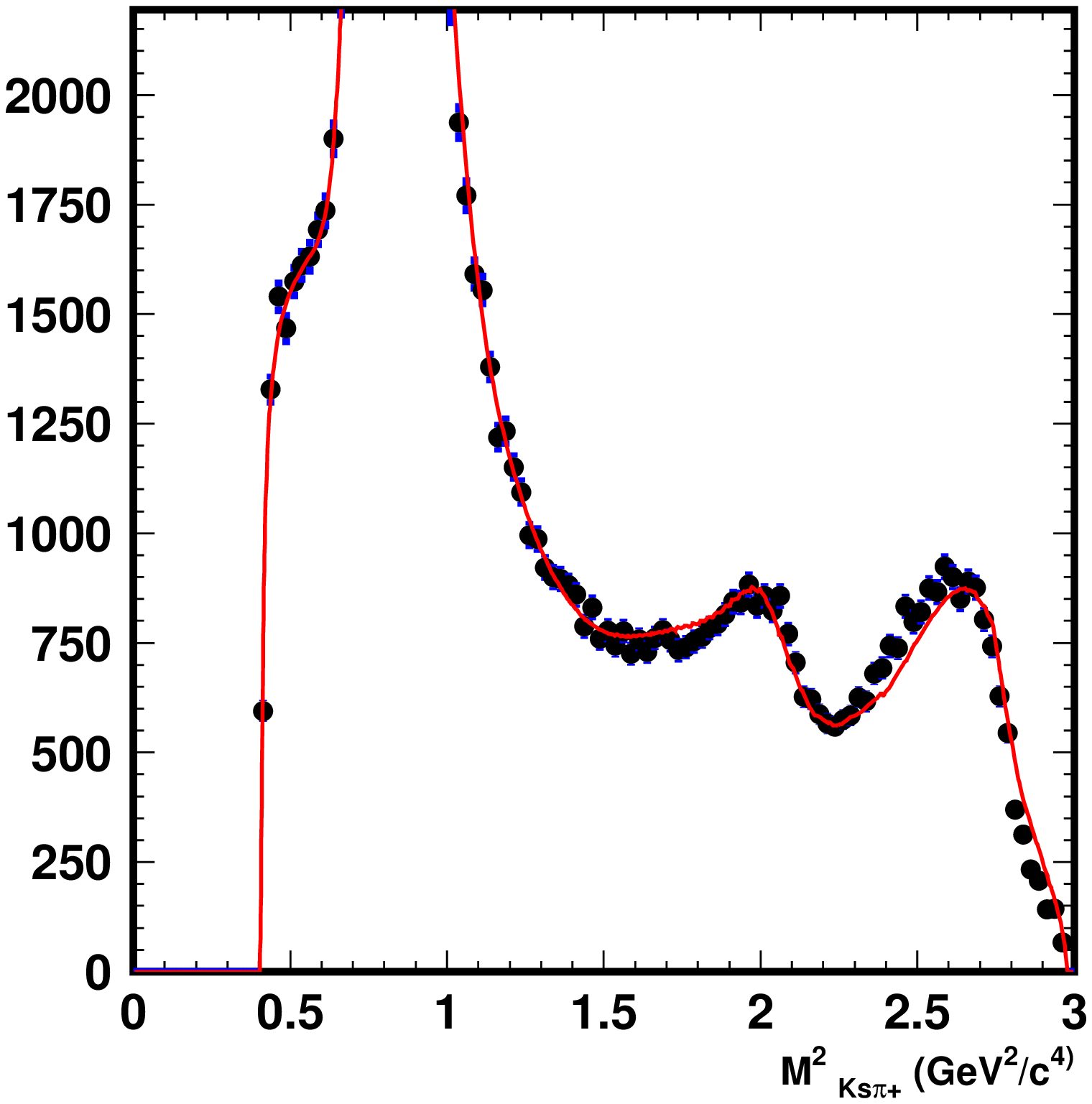}
    \epsfig{width=0.23\textwidth,file=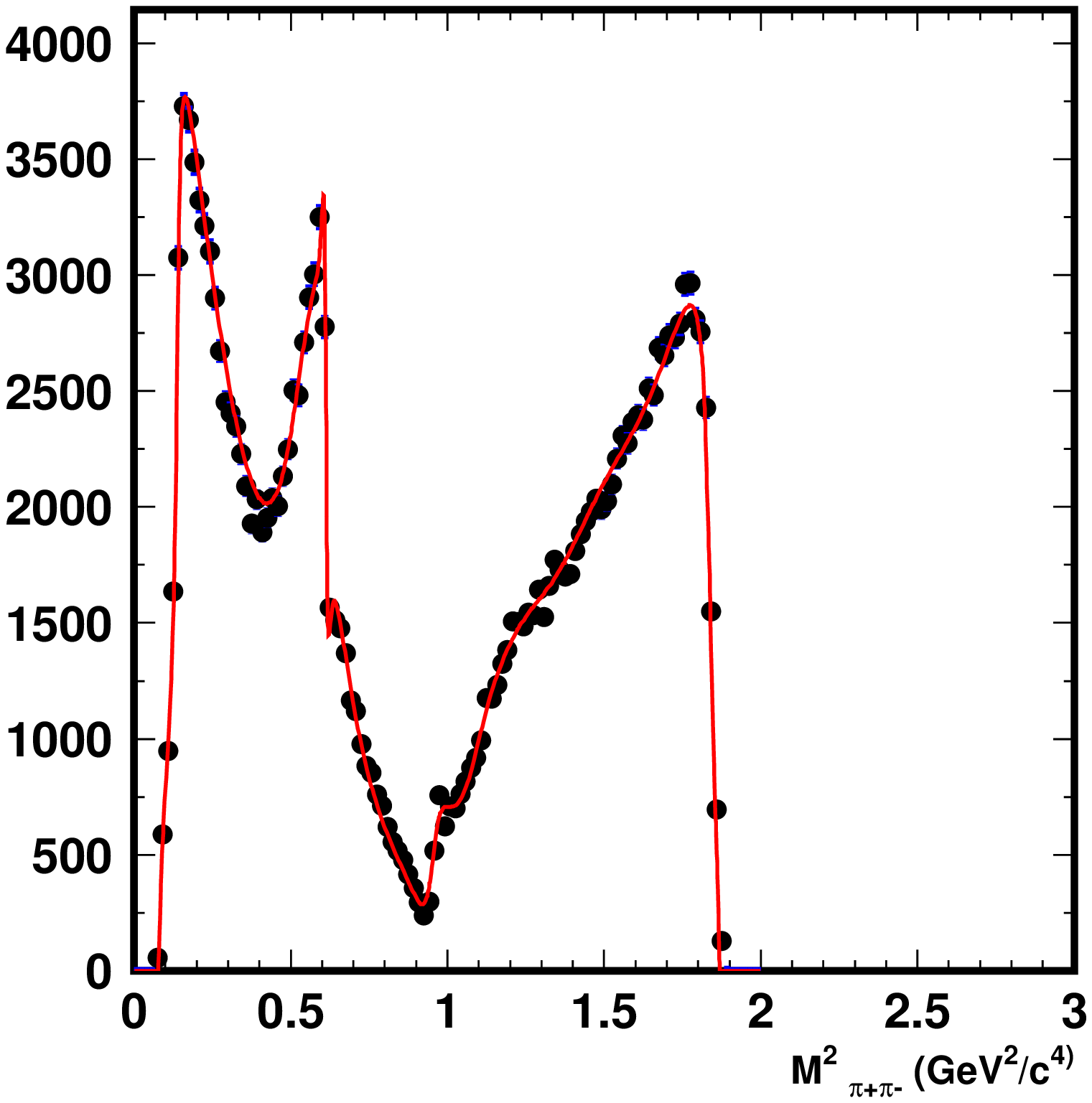}
    \caption{
      From left to right:
      Dalitz plot distribution of tagged $D$ decays;
      projections onto the $m_-^2$, $m_+^2$ and $m_{\pi^+\pi^-}^2$ variables.
    }
    \label{fig:model}
  \end{center}
\end{figure}

We select candidates for the $D \to K_S \pi^+\pi^-$ decay,
and combine with slow charged pions ($\pi_s^\pm$) to make $D^{*\pm}$ candidates.
We require $D^{*\pm}$ candidates to have momenta above 
$2.7 \ {\rm GeV}/c^2$ in the centre of mass (cm) frame 
to reject background from $B$ decay.
We fit the mass difference $\Delta M = M_{K_S \pi^+\pi^- \pi_s^\pm} - M_{K_S \pi^+\pi^-}$
to obtain signal and background fractions, 
and then require $144.6 \ {\rm MeV}/c^2 < \Delta M < 146.4 \ {\rm MeV}/c^2$. 
The distribution of background events across the Dalitz plot
is obtained from a sideband region of $\Delta M$.
We then fit the candidates to a model including 
the components listed in Table.~\ref{tab:model},
with efficiency, resolution and background taken into account.
The result of the fit is shown in Fig.~\ref{fig:model}.
The $\chi^2/{\rm ndf}$ of the fit is $2543/1106$, 
however fine tuning the model has little effect on $\phi_3$
and possible discrepancies are taken into account in the model uncertainty.

\begin{table}[htbp]
  \begin{center}
    \scalebox{0.88}{
    \begin{tabular}{|l|ccc|} 
      \hline
      Resonance & Amplitude & Phase  ($^{\circ}$) & Fraction \\
      \hline
$K_S \sigma_1$               & $1.57   \pm 0.10  $ & $214\pm 4$  & 9.8\%   \\
$K_S \rho^0$               & $1.0$ (fixed)     & 0 (fixed) & 21.6\%  \\
$K_S \omega$                 & $0.0310 \pm 0.0010$ & $113.4 \pm 1.9$ &  0.4\% \\
$K_S f_0(980)$          & $0.394  \pm 0.006 $ & $207\pm 3$      &  4.9\% \\
$K_S \sigma_2$               & $0.23   \pm 0.03  $ & $210\pm 13$     &  0.6\% \\
$K_S f_2(1270)$         & $1.32   \pm 0.04  $ & $348\pm 2$      &  1.5\% \\
$K_S f_0(1370)$         & $1.25   \pm 0.10  $ & $69\pm 8$       &  1.1\% \\
$K_S \rho^0(1450)$         & $0.89   \pm 0.07  $ & $1\pm 6$        &  0.4\% \\
$K^*(892)^+ \pi^-$        & $1.621  \pm 0.010 $ & $131.7\pm 0.5$  & 61.2\% \\
$K^*(892)^- \pi^+$        & $0.154  \pm 0.005 $ & $317.7\pm 1.6$  &  0.55\% \\
$K^*(1410)^+ \pi^-$       & $0.22   \pm 0.04  $ & $120\pm 14$     &  0.05\% \\
$K^*(1410)^- \pi^+$       & $0.35   \pm 0.04  $ & $253\pm 6$      &  0.14\% \\
$K_0^*(1430)^+ \pi^-$     & $2.15   \pm 0.04  $ & $348.7\pm 1.1$  &  7.4\%  \\
$K_0^*(1430)^- \pi^+$     & $0.52   \pm 0.04  $ & $89\pm 4$       &  0.43\% \\
$K_2^*(1430)^+ \pi^-$     & $1.11   \pm 0.03  $ & $320.5\pm 1.8$  &  2.2\%  \\
$K_2^*(1430)^- \pi^+$     & $0.23   \pm 0.02  $ & $263\pm 7$      &  0.09\% \\
$K^*(1680)^+ \pi^-$       & $2.34   \pm 0.26  $ & $110\pm 5$      &  0.36\% \\
$K^*(1680)^- \pi^+$       & $1.3    \pm 0.2   $ & $87\pm 11$      &  0.11\% \\
nonresonant             & $3.8    \pm 0.3   $ & $157\pm 4$      &  9.7\%  \\
      \hline
    \end{tabular}
    }
    \caption{
      Results of the fit to determine the $D$ decay model.
    }
    \label{tab:model}
  \end{center}    
\end{table}

\subsection{Selection of $B$ Decay Candidates}

\begin{figure}[htb]
  \begin{center}
    \epsfig{width=0.33\textwidth,file=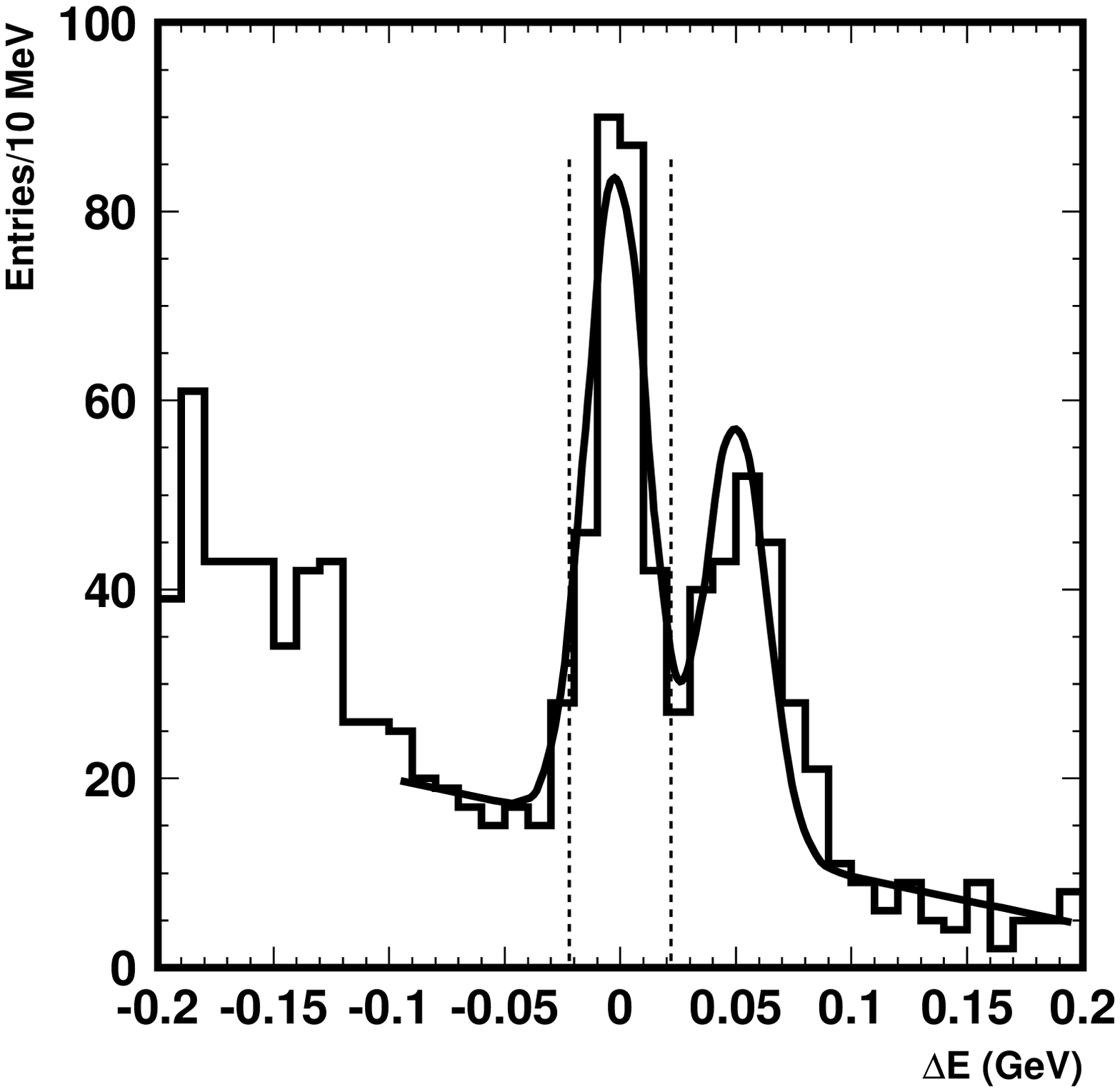} 
    \hspace{0.15\textwidth}
    \epsfig{width=0.33\textwidth,file=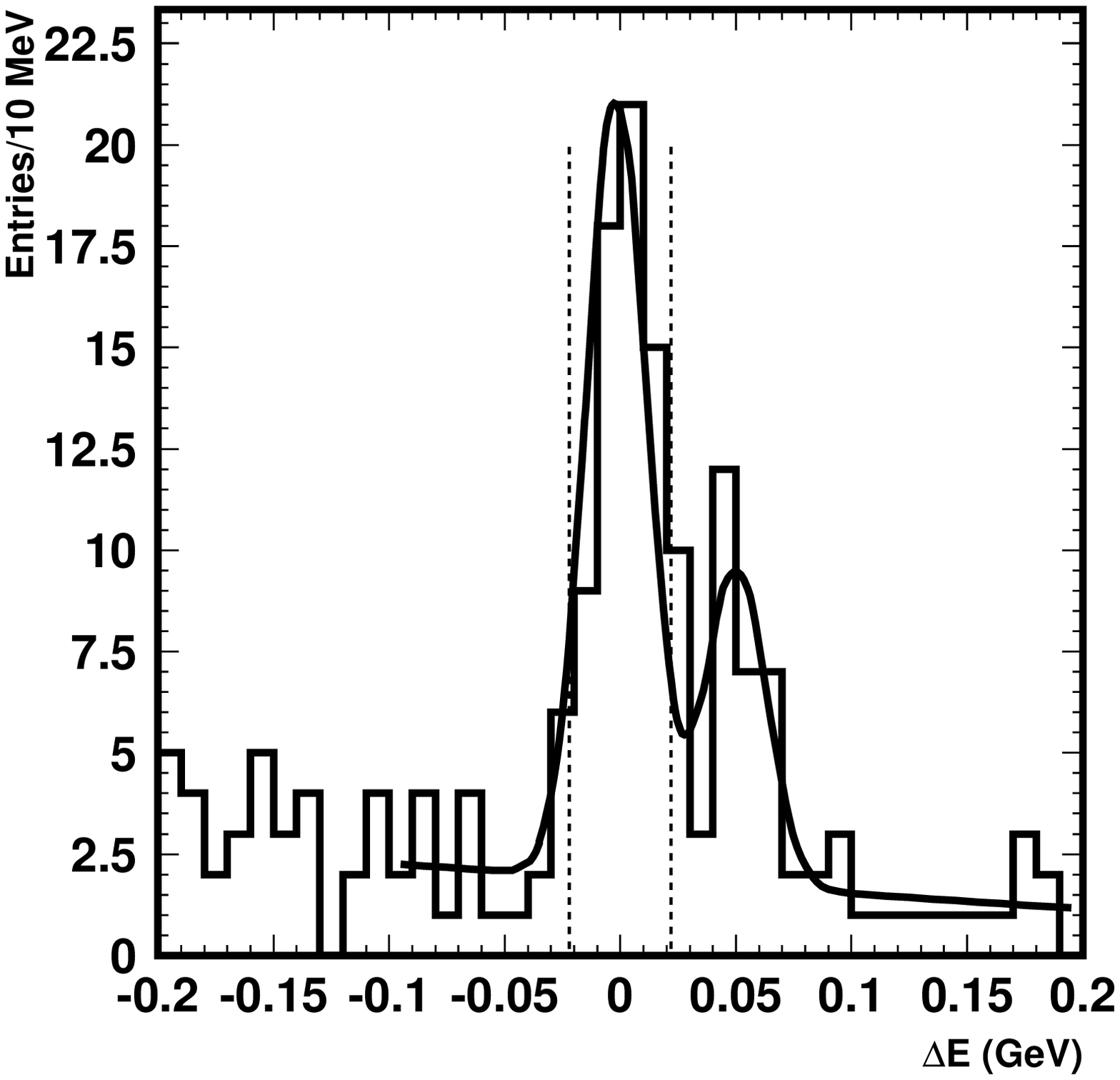} \\
    \epsfig{width=0.24\textwidth,file=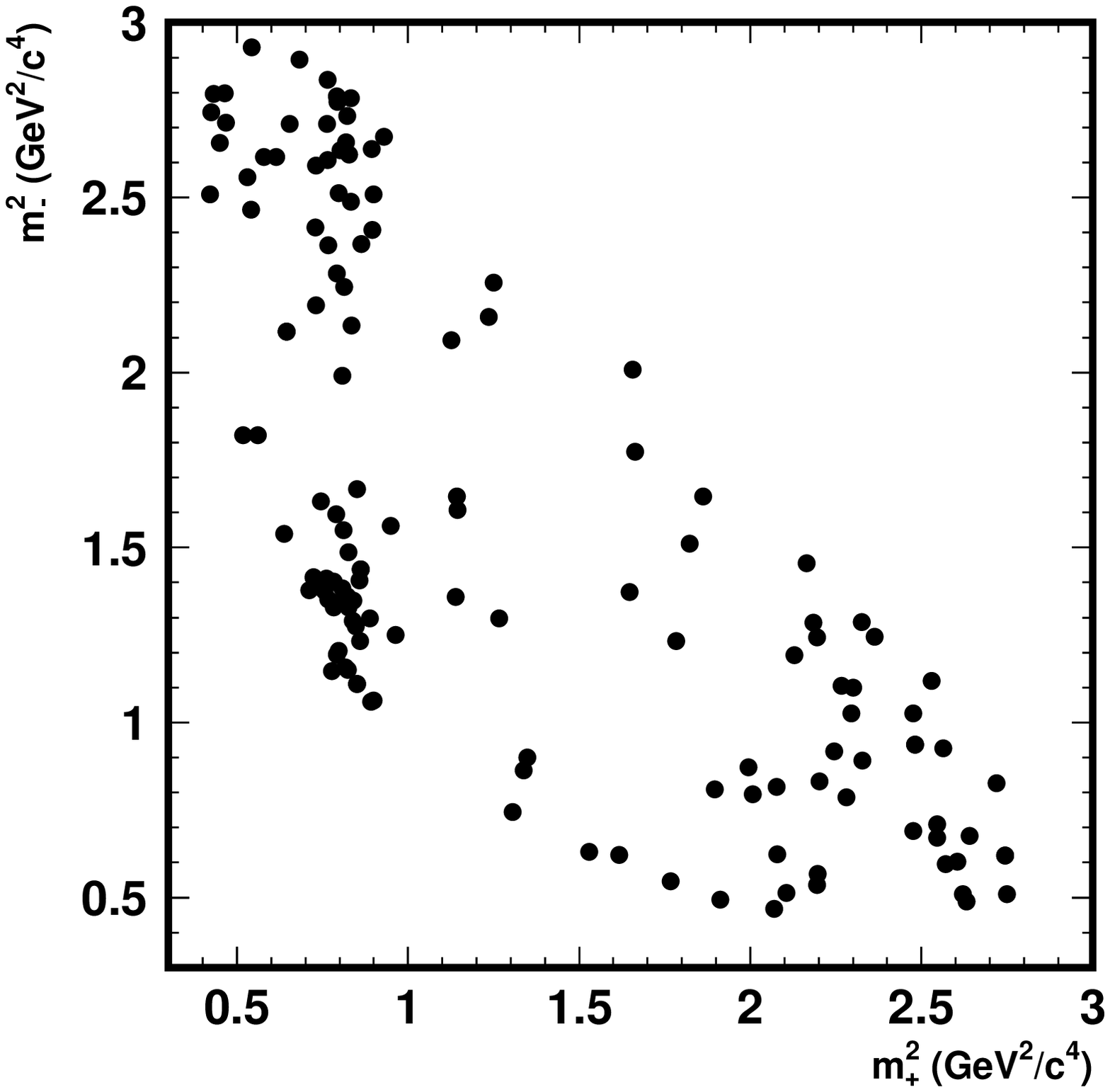} 
    \epsfig{width=0.24\textwidth,file=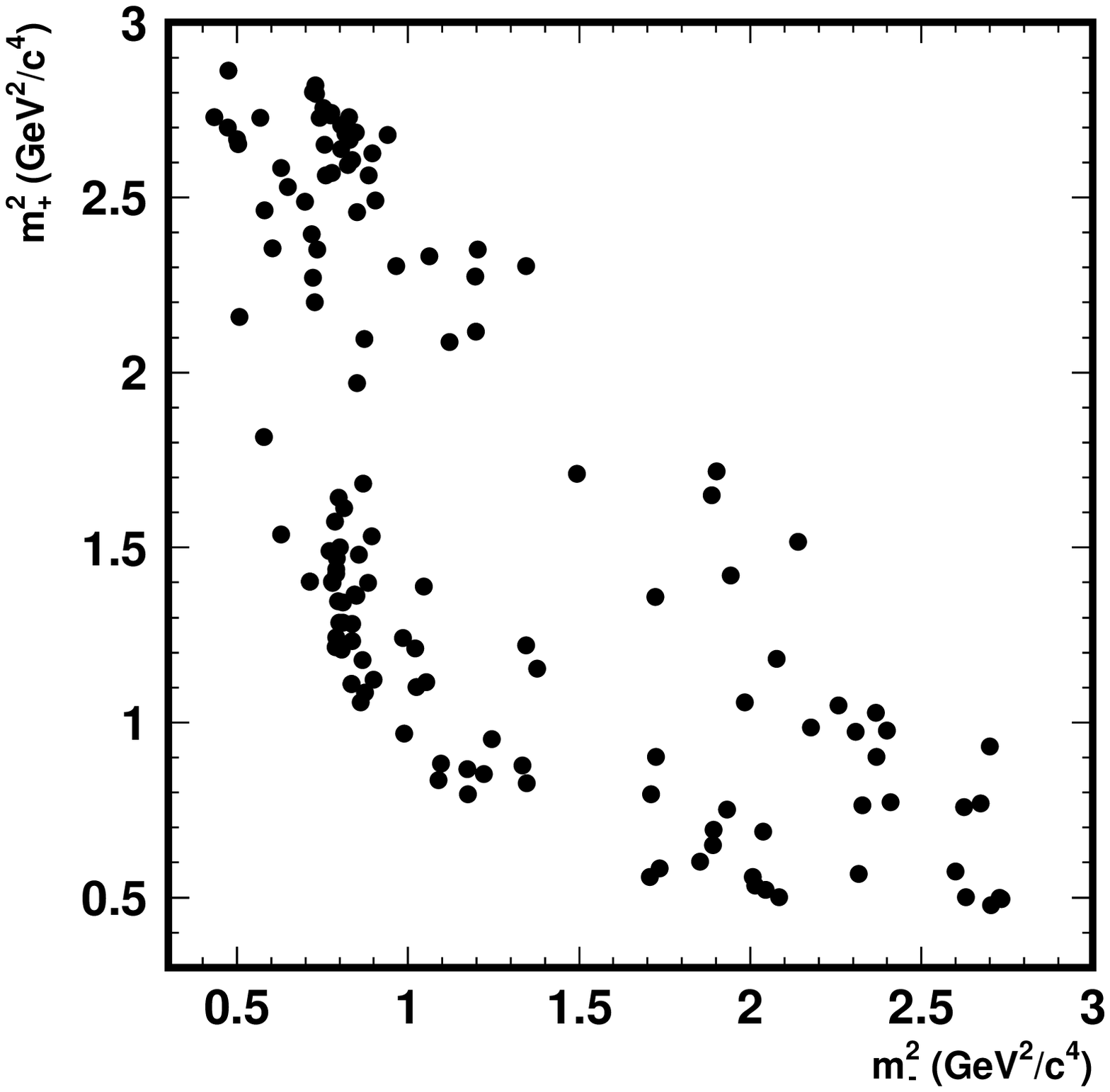} 
    \epsfig{width=0.24\textwidth,file=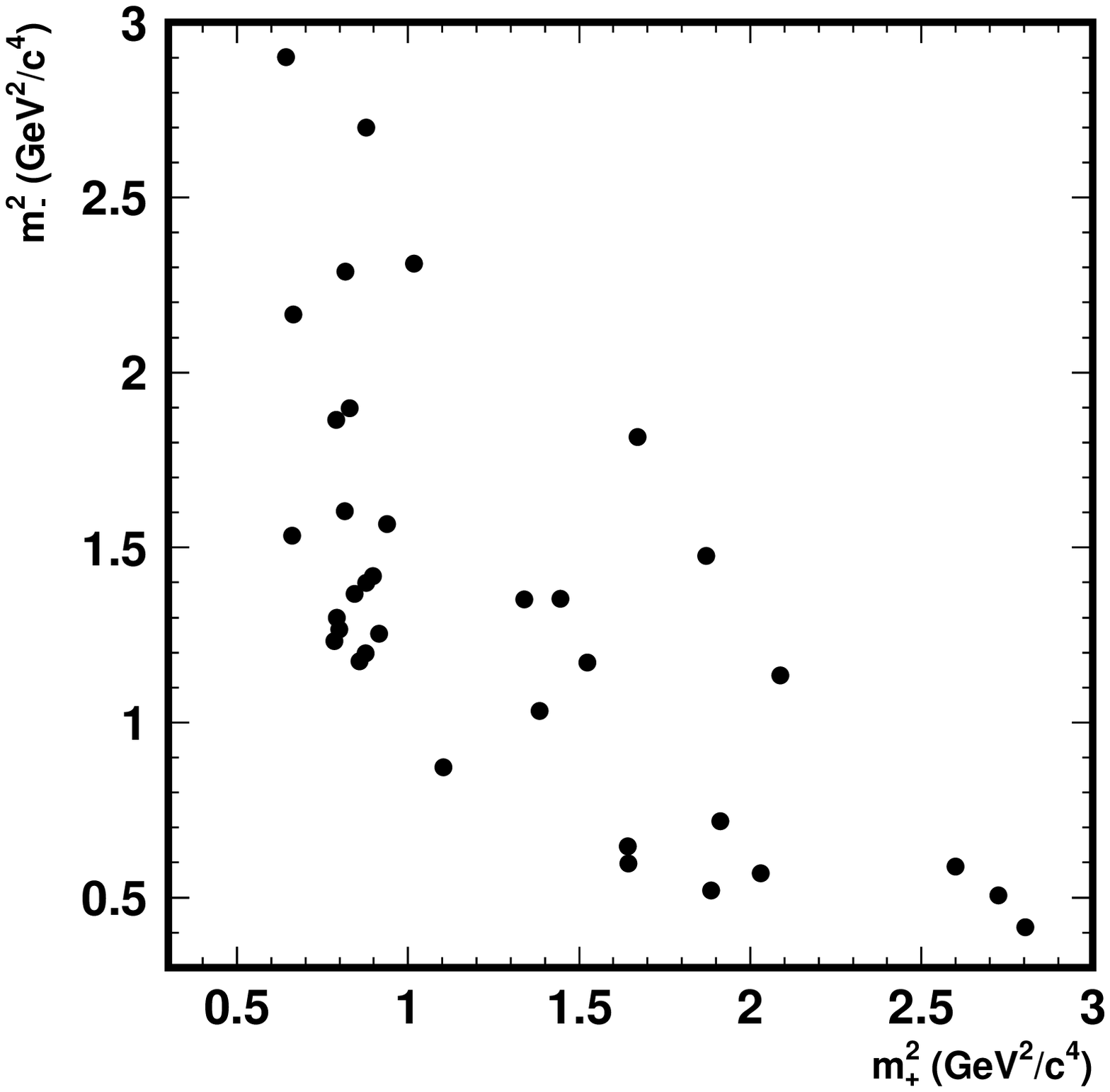} 
    \epsfig{width=0.24\textwidth,file=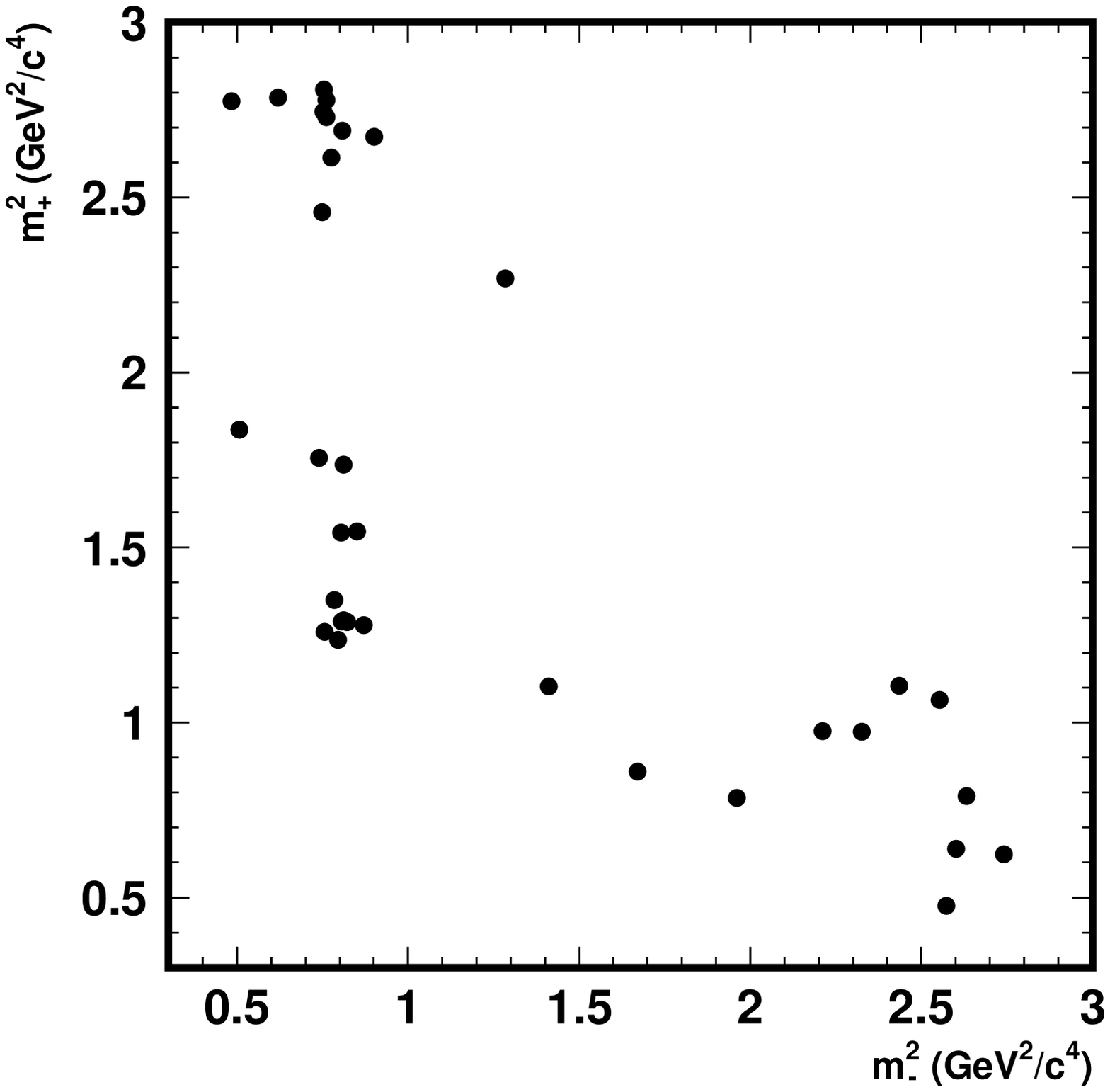}     
    \caption{
      (Top)
      $\Delta E$ distributions and fit results for 
      (left) $B^\pm \to DK^\pm$ and (right) $B^\pm \to D^*K^\pm$ candidates.
      (Bottom) 
      Dalitz plots for (from left to right)
      $B^+ \to DK^+$, $B^- \to DK^-$, 
      $B^+ \to D^*K^+$ and $B^- \to D^*K^-$ candidates.
      The axes are exchanged between $B^+$ and $B^-$ candidates
      so that $CP$ violation effects appear as differences between
      the $B^+$ and $B^-$ distributions.
    }
    \label{fig:de}
  \end{center}
\end{figure}

We reconstruct neutral $D$ mesons in the $D \to K_S \pi^+\pi^-$ mode,
and $D^*$ mesons by making combinations with $\pi^0$ candidates.
We then combine these with prompt tracks,
which are selected as kaon candidates
on the basis of information from the CDC, TOF and ACC systems.
The selection of $B$ candidates is based on the CM energy difference
$\Delta E = \sum E_i - E_{\rm beam}$ and the beam-constrained $B$ meson mass
$M_{\rm bc} = \sqrt{E_{\rm beam}^2 - (\sum p_i)^2}$, 
where $E_{\rm beam}$ is the CM beam energy, 
and $E_i$ and $p_i$ are the CM energies and momenta of the
$B$ candidate decay products. 
The requirements for signal candidates are 
$5.272 \ {\rm GeV}/c^2 < M_{\rm bc} < 5.288 \ {\rm GeV}/c^2$ and 
$|\Delta E| < 0.022 \ {\rm GeV}$.
The signal and background fractions are obtained from binned fits to
the $\Delta E$ distributions, which are shown in Fig.~\ref{fig:de}.
The signal is modelled with a Gaussian,
an additional Gaussian is used to describe the background from
$B^\pm \to D^{(*)} \pi^\pm$ with the prompt pion misidentified as a kaon,
that peaks at $\Delta E \sim 50 \ {\rm MeV}$.
The remaining background is modelled by a linear function.
For $B^\pm \to DK^\pm$ the selection efficiency is $11\%$,
there are 276 candidates satisfying all selection criteria of
which $209 \pm 16$ are found to be signal events.
For $B^\pm \to D^*K^\pm$ the selection efficiency is $6.2\%$,
there are 69 candidates satisfying all selection criteria of
which $58\pm 8$ are found to be signal events.
The Dalitz plot distributions of candidate events 
are also shown in Fig.~\ref{fig:de}.
The Dalitz plot distributions of the different background sources
are obtained using sideband regions in data or Monte Carlo simulation.

\subsection{Extraction of $\phi_3$}

In order to test the analysis procedure, 
we use control samples of $B^\pm \to D^{(*)}\pi^\pm$;
these have similar topology and phenomenology to our signal,
but in this case the ratio of amplitudes is expected to be small ($r \sim 0.01$).
We obtain results which are consistent with expectation,
and take possible discrepancies into account in the systematic error.

\begin{figure}[htb]
  \begin{center}
    \epsfig{width=0.23\textwidth,file=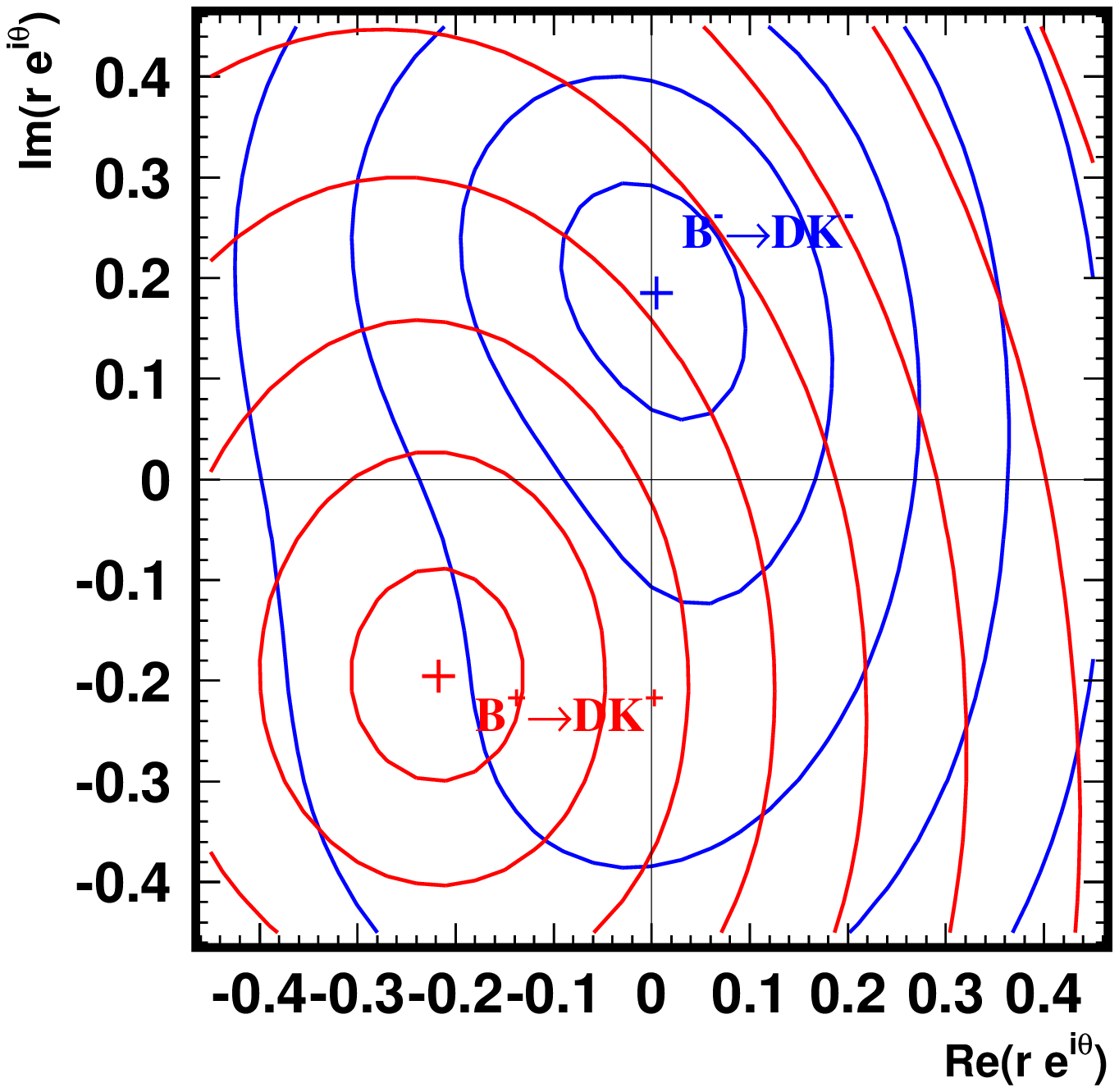} 
    \epsfig{width=0.23\textwidth,file=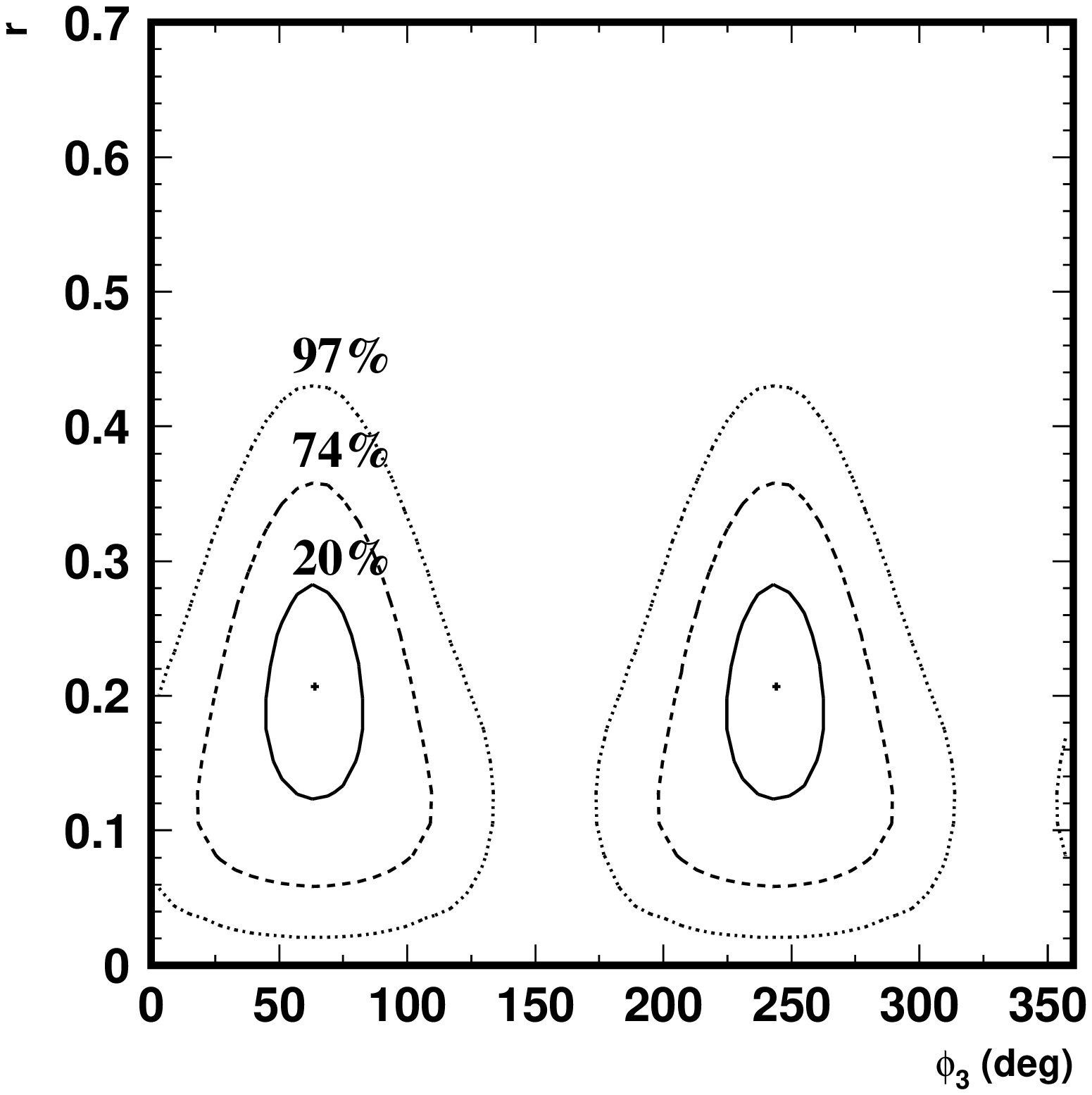}
    \hspace{0.05\textwidth}
    \epsfig{width=0.23\textwidth,file=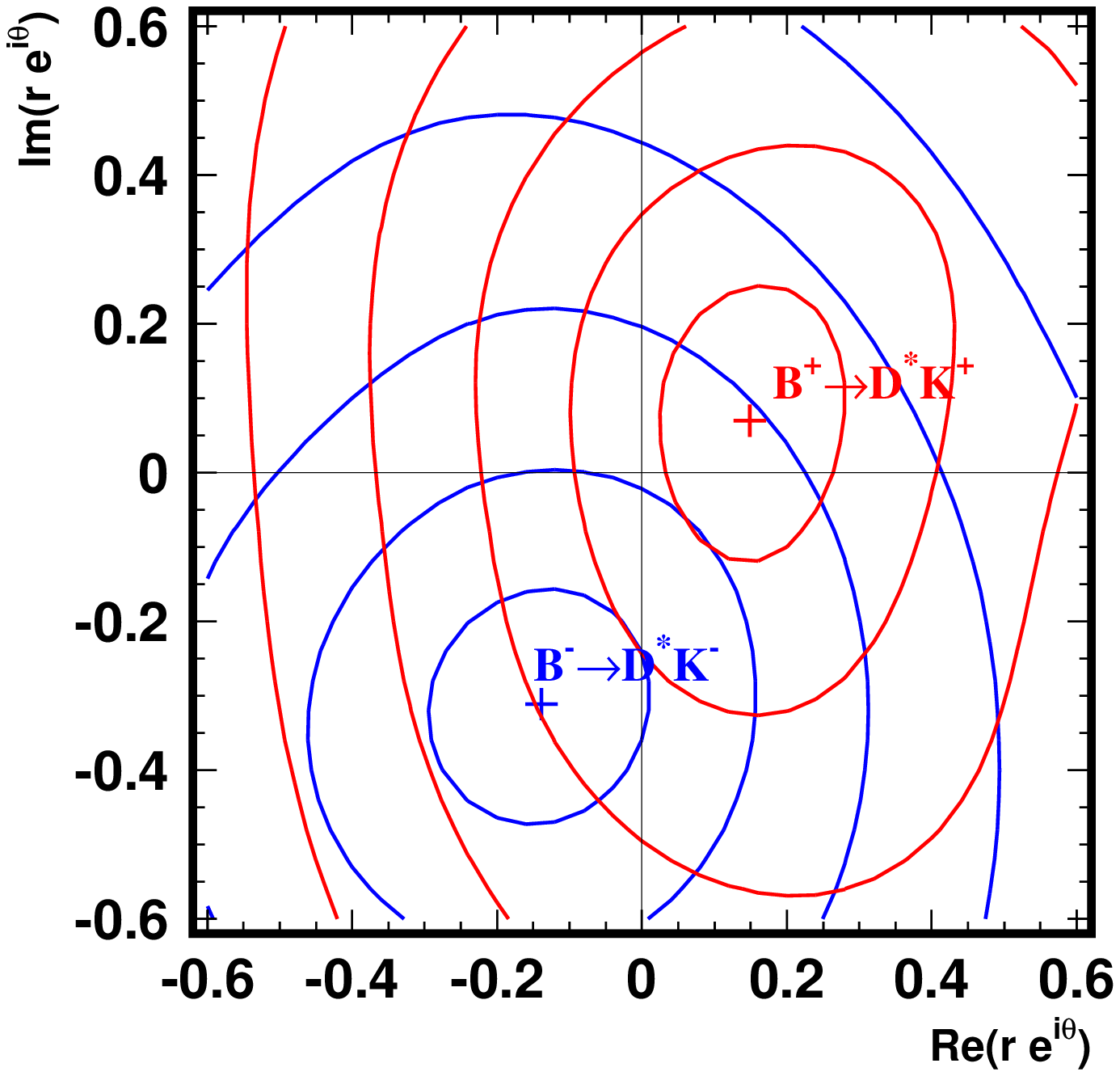} 
    \epsfig{width=0.23\textwidth,file=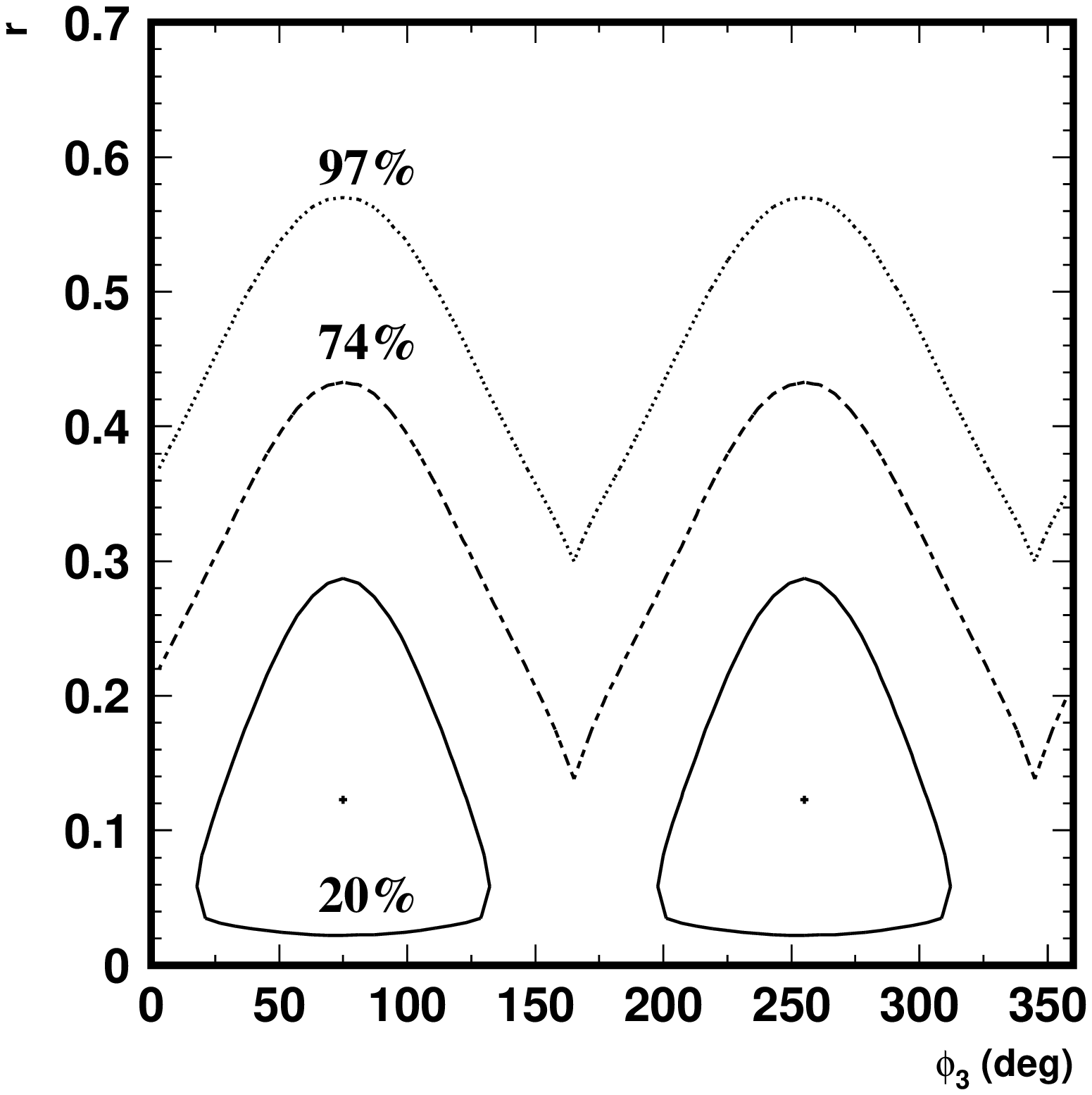}     
    \caption{
      Likelihood contours for
      (left) $B^\pm \to DK^\pm$ and (right) $B^\pm \to D^*K^\pm$ candidates.
      In each pair, the leftmost plot is the result of the unbinned fit,
      and the rightmost is the outcome of the frequentist treatment,
      as described in the text.
    }
    \label{fig:results}
  \end{center}
\end{figure}

We then perform unbinned maximum likelihood fits to the
$B^\pm \to D^{(*)}K^\pm$ samples.
In order to extract the three free parameters $\phi_3$, $r$ and $\delta$,
we fit the $B^+$ and $B^-$ samples simultaneously.
However, to display the result, as in Fig.~\ref{fig:results}, 
we show likelihood contours
in $({\rm Re}(re^{i\theta_\pm}), {\rm Im}(re^{i\theta_\pm}))$,
that are obtained separately for $B^+$ and $B^-$.
$CP$ violation is seen as a difference between the $B^+$ and $B^-$ contours.

Since the value of $r$ is positive definite,
and the uncertainty on $\phi_3$ depends on $r$,
we do not rely on the errors obtained from the 
likelihood curves~\cite{fit_results}, 
but instead use a frequentist approach to obtain confidence regions,
which are also shown in Fig.~\ref{fig:results}.
The results for $B^\pm \to DK^\pm$ are 
\begin{eqnarray}
  \phi_3 & = & 64^\circ \pm 19^\circ \stat \pm 13^\circ \syst \pm 11^\circ \model, \nonumber \\
  r   & = & 0.21 \pm 0.08 \stat \pm 0.03 \syst \pm 0.04 \model, \\
  \delta   & = & 157^\circ \pm 19^\circ \stat \pm 11^\circ \syst \pm 21^\circ \model, \nonumber
\end{eqnarray}
which correspond to $r > 0$ at $99.3\% \ {\rm CL}$ 
and $CP$ violation at $94\% \ {\rm CL}$.
The results for $B^\pm \to D^*K^\pm$ are
\begin{eqnarray}
  \phi_3 & = & 75^\circ \pm 57^\circ \stat \pm 11^\circ \syst \pm 11^\circ \model, \nonumber \\
  r   & = & 0.12 \,^{+0.16}_{-0.11} \stat \pm 0.02 \syst \pm 0.04 \model, \\
  \delta   & = & 321^\circ \pm 57^\circ \stat \pm 11^\circ \syst \pm 21^\circ \model, \nonumber
\end{eqnarray}
 which correspond to $r > 0$ at $56\% \ {\rm CL}$
and $CP$ violation at $38\% \ {\rm CL}$.
The dominant systematic uncertainty is due to possible fit bias
(note again that more details can be found in~\cite{belle_new}).

To combine the results from both modes
we use a frequentist approach with Feldman-Cousins ordering 
and obtain
\begin{equation}
  \phi_3 = 68^\circ \,^{+14^\circ}_{-15^\circ} \stat \pm 13^\circ \syst \pm 11^\circ \model,
\end{equation}
which corresponds to $CP$ violation at $98\% \ {\rm CL}$.
The two standard deviation interval 
including systematic and model uncertainties is $22^\circ < \phi_3 < 113^\circ$.

\section{Summary}

The most precise measurement of $\phi_3$ is presented.
The value is consistent with the previous measurements 
from Belle~\cite{belle_phi3} and BaBar~\cite{babar_phi3}.
We obtain a significantly smaller statistical error than BaBar
due to the larger value of $r$ obtained
(there is, however, no significant discrepancy in the values of $r$
obtained from these and other measurements~\cite{exp_ads}).

\section*{Acknowledgements}

The author thanks the organizers for a highly enjoyable conference.
The author is supported by the Japan Society for the Promotion of Science.

\label{GershonEnd}

\end{document}

%% file: fpcp04-macro.tex
\def\beq{\begin{equation}}
\def\eeq#1{\label{#1}\end{equation}}
\def\eeqn{\end{equation}}

\def\beqa{\begin{eqnarray}}
\def\eeqa#1{\label{#1}\end{eqnarray}}
\def\eeqan{\end{eqnarray}}

\let\bar=\overbar

\def\Dslash{\not{\hbox{\kern-4pt $D$}}}
\def\dslash{\not{\hbox{\kern-2pt $\del$}}}

\def\msb{{\bar{\ssstyle M \kern -1pt S}}}

\def\BB0bar{B^0 {\overline B}^0}
\def\BB0dbar{B_d^0 {\overline B}_d^0}
\def\BB0sbar{B_s^0 {\overline B}_s^0}

\RequirePackage{xspace}

\usepackage{relsize}
\def\babar{\mbox{\slshape B\kern-0.1em{\smaller A}\kern-0.1em
    B\kern-0.1em{\smaller A\kern-0.2em R}}}

\def\Kbar  {\kern 0.2em\overline{\kern -0.2em K}{}\xspace}

\def\Kz    {\ensuremath{K^0}\xspace}
\def\Kzb   {\ensuremath{\Kbar^0}\xspace}
\def\KzKzb {\ensuremath{\Kz \kern -0.16em \Kzb}\xspace}
\def\Kp    {\ensuremath{K^+}\xspace}
\def\Km    {\ensuremath{K^-}\xspace}

\def\KpKm  {\ensuremath{\Kp \kern -0.16em \Km}\xspace}

\def\Dbar    {\kern 0.2em\overline{\kern -0.2em D}{}\xspace}

\def\Dz      {\ensuremath{D^0}\xspace}
\def\Dzb     {\ensuremath{\Dbar^0}\xspace}
\def\DzDzb   {\ensuremath{\Dz {\kern -0.16em \Dzb}}\xspace}
\def\Dp      {\ensuremath{D^+}\xspace}
\def\Dm      {\ensuremath{D^-}\xspace}

\def\DpDm    {\ensuremath{\Dp {\kern -0.16em \Dm}}\xspace}

\def\Bbar    {\kern 0.18em\overline{\kern -0.18em B}{}\xspace}

\def\BB      {\ensuremath{B\Bbar}\xspace} 
\def\Bz      {\ensuremath{B^0}\xspace}
\def\Bzb     {\ensuremath{\Bbar^0}\xspace}
\def\BzBzb   {\ensuremath{\Bz {\kern -0.16em \Bzb}}\xspace}
\def\Bu      {\ensuremath{B^+}\xspace}
\def\Bub     {\ensuremath{B^-}\xspace}

\def\BpBm    {\ensuremath{\Bu {\kern -0.16em \Bub}}\xspace}

\mathchardef\Upsilon="7107
\def\Y#1S{\ensuremath{\Upsilon{(#1S)}}\xspace}

\mathchardef\Deltares="7101
\mathchardef\Xi="7104
\mathchardef\Lambda="7103
\mathchardef\Sigma="7106
\mathchardef\Omega="710A

\def\Deltabar{\kern 0.25em\overline{\kern -0.25em \Deltares}{}\xspace}
\def\Lbar{\kern 0.2em\overline{\kern -0.2em\Lambda\kern 0.05em}\kern-0.05em{}\xspace}
\def\Sigbar{\kern 0.2em\overline{\kern -0.2em \Sigma}{}\xspace}
\def\Xibar{\kern 0.2em\overline{\kern -0.2em \Xi}{}\xspace}
\def\Obar{\kern 0.2em\overline{\kern -0.2em \Omega}{}\xspace}
\def\Nbar{\kern 0.2em\overline{\kern -0.2em N}{}\xspace}
\def\Xb{\kern 0.2em\overline{\kern -0.2em X}{}\xspace}

\newcommand{\tev}{\ensuremath{\mathrm{\,Te\kern -0.1em V}}\xspace}
\newcommand{\gev}{\ensuremath{\mathrm{\,Ge\kern -0.1em V}}\xspace}
\newcommand{\mev}{\ensuremath{\mathrm{\,Me\kern -0.1em V}}\xspace}
\newcommand{\kev}{\ensuremath{\mathrm{\,ke\kern -0.1em V}}\xspace}
\newcommand{\ev}{\ensuremath{\mathrm{\,e\kern -0.1em V}}\xspace}
\newcommand{\gevc}{\ensuremath{{\mathrm{\,Ge\kern -0.1em V\!/}c}}\xspace}
\newcommand{\mevc}{\ensuremath{{\mathrm{\,Me\kern -0.1em V\!/}c}}\xspace}
\newcommand{\gevcc}{\ensuremath{{\mathrm{\,Ge\kern -0.1em V\!/}c^2}}\xspace}
\newcommand{\mevcc}{\ensuremath{{\mathrm{\,Me\kern -0.1em V\!/}c^2}}\xspace}

\def\mus  {\ensuremath{\rm \,\mus}\xspace}

\def\mus        {\ensuremath{\,\mu{\rm s}}\xspace}    

\def\to                 {\ensuremath{\rightarrow}\xspace}

\newcommand{\stat}{\ensuremath{\mathrm{(stat)}}\xspace}
\newcommand{\syst}{\ensuremath{\mathrm{(syst)}}\xspace}

\def\pep2{PEP-II}

\def\gsim{{~\raise.15em\hbox{$>$}\kern-.85em
          \lower.35em\hbox{$\sim$}~}\xspace}
\def\lsim{{~\raise.15em\hbox{$<$}\kern-.85em
          \lower.35em\hbox{$\sim$}~}\xspace}

\xspace

\def\jetset74   {\mbox{\tt Jetset \hspace{-0.5em}7.\hspace{-0.2em}4}\xspace}


%% file: gershon_sub.bbl
\begin{thebibliography}{99}

\bibitem{bevan}
  A.~Bevan, these proceedings.
  K.~Abe {\it et al.} (Belle Collaboration),
  Phys. Rev. Lett. {\bf 93}, 021601 (2004).
\bibitem{graziani}
  G.~Graziani, these proceedings.
  Y.~Chao {\it et al.} (Belle Collaboration), 
  Phys. Rev. Lett. {\bf 93}, 191802 (2004).
\bibitem{belle_pi0pi0}
  K.~Abe {\it et al.} (Belle Collaboration), 
  BELLE-CONF-0406 (hep-ex/0408101).
\bibitem{belle_rho0pi0}
  J.~Dragic, T.~Gershon {\it et al} (Belle Collaboration), 
  Phys. Rev. Lett. {\bf 93}, 131802 (2004).
\bibitem{skchoi}
  S.K.~Choi, these proceedings.
\bibitem{belle:btos}
  K.~Abe {\it et al.} (Belle Collaboration), 
  BELLE-CONF-0435 (hep-ex/0409049).
\bibitem{hazumi}
  M.~Hazumi, these proceedings.
  K.~Abe {\it et al.} (Belle Collaboration), 
  BELLE-CONF-0475 (hep-ex/0411056).
\bibitem{belle_conf}
  Please see http://belle.kek.jp/conferences/ICHEP2004/ 
  for a complete list.
\bibitem{yjkwon}
  Y.J.~Kwon, these proceedings.
  K.~Abe {\it et al.} (Belle Collaboration), 
  BELLE-CONF-0415 (hep-ex/0410006).
\bibitem{kekb}
  S.~Kurokawa and E.~Kikutani, 
  Nucl. Instr. and. Meth. A {\bf 499}, 1 (2003),
  and other papers included in this volume.
\bibitem{belle}
  A.~Abashian {\it et al.} (Belle collaboration), 
  Nucl. Instr. and Meth. A {\bf 479}, 117 (2002).
\bibitem{ushiroda} 
  Y.~Ushiroda (Belle SVD2 Group), 
  Nucl. Instr. and Meth. A {\bf 511}, 6 (2003).
\bibitem{ckm}
  M.~Kobayashi and T.~Maskawa, Prog. Theor. Phys. {\bf 49}, 652 (1973); 
  N.~Cabibbo, Phys. Rev. Lett. {\bf 10}, 531 (1963).
\bibitem{bigisanda}
  I.I.~Bigi and A.I.~Sanda, Phys. Lett. B {\bf 211}, 213 (1988).
\bibitem{phi3_methods}
  M.~Gronau and D.~London, Phys. Lett. B {\bf 253}, 483 (1991);
  M.~Gronau and D.~Wyler, Phys. Lett. B {\bf 265}, 172 (1991).
  I.~Dunietz, Phys. Lett. {\bf B270}, 75 (1991).
  D.~Atwood, G.~Eilam, M.~Gronau and A.~Soni, 
  Phys. Lett. B {\bf 341}, 372 (1995).
  D.~Atwood, I.~Dunietz and A.~Soni, 
  Phys. Rev. Lett. {\bf 78}, 3257 (1997);
  Phys. Rev. D {\bf 63}, 036005 (2001).
\bibitem{ggsz}
  A.~Giri, Yu.~Grossman, A.~Soffer, J.~Zupan, 
  Phys. Rev. D {\bf 68}, 054018 (2003).\\
  This technique was proposed independently in the Belle Collaboration, 
  and the analysis of experimental data was under way before the 
  A.~Giri {\it et al.} publication appeared
  (Proceedings of BINP Special Analysis Meeting on Dalitz Analysis, 
  24-26 Sep. 2002, unpublished). 
\bibitem{zupan}
  J.~Zupan, these proceedings.
\bibitem{bg}
  A.~Bondar and T.~Gershon,
  Phys. Rev. D {\bf 70}, 091503(R) (2004).
\bibitem{belle_phi3}
  A.~Poluektov {\it et al.} (Belle Collaboration), 
  Phys. Rev. D {\bf 70}, 072003 (2004).
\bibitem{babar_phi3}
  B.~Aubert {\it et al.} (BABAR Collaboration),
  BABAR-CONF-04/043 (hep-ex/0408088);
  See also D.~Lange, these proceedings.
\bibitem{belle_new}
  K.~Abe {\it et al.} (Belle Collaboration), 
  BELLE-CONF-0476 (hep-ex/0411049).
\bibitem{fit_results}
  The results of the fit to $B^\pm \to DK^\pm$ candidates are
  $r = 0.247 \pm 0.071$, $\phi_3 = 63.7^\circ \pm 15.2^\circ$ and $\delta = 156.6^\circ \pm 15.6^\circ$;
  the results of the fit to $B^\pm \to D^*K^\pm$ candidates are
  $r = 0.254 \pm 0.116$, $\phi_3 = 74.9^\circ \pm 25.2^\circ$ and $\delta = 321.3^\circ \pm 25.0^\circ$.
\bibitem{exp_ads}
  K.~Abe {\it et al.} (Belle Collaboration), 
  BELLE-CONF-0444 (hep-ex/0408129);
  B.~Aubert {\it et al.} (BABAR Collaboration),
  BABAR-CONF-04/013 (hep-ex/0408028);
  

\end{thebibliography}
